\documentstyle[epsfig]{aipproc}
\begin{document}

\title{Temperature-dependent dielectric and piezoelectric response
of ferroelectrics from first principles\thanks{To be published in the
proceedings of the Fifth Williamsburg Workshop on First-Principles Calculations
for Ferroelectrics, February 1998.}}
\author{K. M. Rabe and E. Cockayne}
\address{Department of Applied Physics, Yale University \\
P. O. Box 208284, New Haven, Connecticut, 06520-8284}
\maketitle

\begin{abstract}
A method for the calculation of the temperature dependence of dielectric and
piezoelectric responses, based on the use of a first-principles effective
Hamiltonian, is described. Results are presented for the ferroelectric
perovskite PbTiO$_3$. While the method includes only the soft-mode
contributions to the responses, it is argued to give a good description of the
divergences or near-divergences of the response functions near the
cubic-tetragonal transition. The expression of the response functions in terms
of correlation functions is used to provide a real-space interpretation of the
responses which clearly distinguishes between PbTiO$_3$ and the related
materials BaTiO$_3$ and KNbO$_3$.
\end{abstract}

\section*{Introduction}

Systems with large dielectric and piezoelectric responses are of interest both
from a fundamental and a technological point of view\cite{transducer}.
Understanding the microsopic origin of a high sensitivity of the polarization
of a ferroelectric oxide to applied field, in the case of the dielectric
constant, or stress-induced changes in strain, in the case of piezoelectricity,
can lead to the optimization of these properties through the appropriate choice
of materials.
Empirically, large responses are observed in stoichiometric compounds such as
PbTiO$_3$ and BaTiO$_3$ for temperatures near the ferroelectric phase
transition \cite{remeika,jjap,linew,btexpt}. In solid solutions, such as PZT,
enhanced responses are observed over a wide temperature range\cite{PZT}.

In this paper, we discuss the modeling of the piezoelectric and dielectric
response of ferroelectric perovskites, specifically PbTiO$_3$ and BaTiO$_3$,
though the use of first-principles effective Hamiltonians previously
constructed in Refs. \onlinecite{rwwannier,ptwburg,ptprb,zvr}.
The calculations give reasonable agreement with experiment, considering that
this modeling gives only the contribution of the soft modes to the responses,
and not other contributions such as thermal expansion and the response of other
polar modes. However, for temperatures near the ferroelectric phase transition,
the soft modes are expected to dominate and thus this approach should capture
the essential physics of the large observed responses.
In addition, it should correctly describe the trends between different
materials. Finally, with this microscopic approach, we are able directly to
relate the calculated response functions to the characteristic correlations of
local unit cell polarizations, which should yield further insight into the
nature of the large responses of interest.

\section*{First-principles effective Hamiltonians}
 The construction of first-principles effective Hamiltonians for PbTiO$_3$ and
BaTiO$_3$ has previously been described in detail \cite{ptwburg,ptprb,zvr}.
Briefly, the model consists of one vector per five-atom unit cell which
represents the local polarization associated with the ferroelectric distortion.
The potential ${\cal H}_{eff}(\{\vec \xi_i\},e_{\alpha\beta})$ for these vector
degrees of freedom is expanded for the high-symmetry cubic perovskite reference
structure, including local anharmonic terms, quadratic intersite interactions
which are assumed to be dipolar beyond third neighbors, and lowest order
coupling to homogeneous strain.

The extension of this model to include the effects of electric field is
accomplished by writing the dependence of the polarization on the model degrees
of freedom to lowest order:
\begin{equation}
P(\{\vec \xi_i\},e_{\alpha\beta})=\sum_i \overline Z^* ea_0\vec \xi_i
\label{poldef}
\end{equation}
and adding to the potential the coupling term
$-\vec P\cdot\vec E$
where $\vec E$ is the macroscopic electric field\cite{Nye}.
The dielectric response is then simply obtained from ${d<P_\alpha>\over
dE_\beta}$, while the piezoelectric response can be expressed as
${d<e_{\alpha\beta}> \over dE_\gamma}$, where the brackets are used to denote
the thermal expectation value.
The effects of macroscopic stress $\sigma_{ij}$ can be included by adding to
the potential the coupling term $-\sigma_{ij} \cdot e_{ij}.$
With these couplings, the thermodynamic identity ${de_{jk} \over dE_{i}}$ =
${dP_i \over d\sigma_{jk}}$ can be obtained (the well-known equality of the
direct and converse piezoelectric effects\cite{Nye}). In the rest of this
paper, we work at zero external stress and evaluate the piezoelectric response
using ${d<e_{\alpha\beta}> \over dE_\gamma}.$

Calculations of the temperature dependent properties of the system are carried
out using classical Monte Carlo calculations, as previously
described\cite{ptwburg,ptprb}. The calculations of the response functions
presented here involved 7x7x7 simulation cells, 10,000 Monte Carlo sweeps (MCS)
for thermalization, 200,000 MCS for the computation of thermal expectation
values, and several runs at each temperature with different random number seeds
to estimate statistical error. The calculations of the real-space correlation
functions presented here involved 10x10x10 simulation cells, 10,000 Monte Carlo
sweeps (MCS) for thermalization, and 100,000 MCS for the computation of thermal
expectation values. In order to compute the temperature-dependence of the order
parameter and thus identify the cubic-tetragonal transition, a small
symmetry-breaking field $E_z$ of magnitude ${20\over\epsilon_\infty}$ kV/cm was
applied. The response function results include this nonzero field.

\section*{Correlation function expressions for response functions}
In the Monte Carlo simulations described above, the response functions are
computed using the fact that they can be expressed as correlation functions.
This allows the appropriate derivatives to be calculated in one Monte Carlo
run.

Specifically, we consider the dielectric tensor:
$$\epsilon_{\alpha\beta}=\epsilon_\infty + 4 \pi  \chi_{\alpha\beta}$$
where
$$ \chi_{\alpha\beta}={\partial <\Omega^{-1}P_{\alpha}> \over \partial
E_{\beta}}.$$
Here, $\Omega$ is the volume of the unit cell and $\vec P$ is the polarization
per unit cell, specifically
\begin{equation}
<P_{i\alpha}>={\int\{de_{\alpha\beta}\}\{\prod_j d\vec \xi_j\}({Z^*ea_0 \over
N}\sum_i \xi_{i\alpha})exp(-\beta(H_{eff}-Z^*ea_0\sum_i \xi_{i\alpha}\cdot\vec
E))\over \int\{de_{\alpha\beta}\}\{\prod_j d\vec
\xi_j\}exp(-\beta(H_{eff}-Z^*ea_0\sum_i \xi_{i\alpha}\cdot\vec E))}
\label{poltherm}
\end{equation}
or
$$<P_{i\alpha}>=({1\over N\beta}){\partial lnZ(\beta,\vec E)\over\partial
E_{\alpha}}$$
where we have defined
$$Z(\beta,\vec E)=\int\{de_{\alpha\beta}\}\{\prod_j d\vec
\xi_j\}exp(-\beta(H_{eff}-Z^*ea_0\sum_i \xi_{i\alpha}\cdot\vec E)).$$
Using the approximation ${\partial <\Omega^{-1}P_{\alpha}> \over \partial
E_{\beta}}=\Omega^{-1}{\partial <P_{\alpha}> \over \partial E_{\beta}}$ (see
footnote 7 in Ref. \onlinecite{wgarcia}) and differentiating Equation
\ref{poltherm} with respect to $E_\alpha$, we readily find the correlation
function expression:
\begin{equation}
\chi_{\alpha\beta}={\beta(Z^*ea_0)^2 \over \Omega}(<\sum_i\xi_{i\alpha}{1\over
N}\sum_j\xi_{j\beta}>-N<\xi_{\alpha}><\xi_{\beta}>)
\label{dicorrfcn}
\end{equation}
where
$$<\xi_{\alpha}>={1\over N}<\sum_i \xi_{i\alpha}>.$$

The piezoelectric tensor can be expressed in terms of correlation functions in
an completely analogous way:
$$ d_{\gamma\alpha\beta}={\partial <e_{\alpha\beta}> \over \partial
E_{\gamma}}$$
where
\begin{equation}
<e_{\alpha\beta}>={\int\{de_{\alpha'\beta'}\}\{\prod_j d\vec
\xi_j\}e_{\alpha\beta}exp(-\beta(H_{eff}-Z^*ea_0\sum_i \vec \xi_{i}\cdot\vec
E))\over \int\{de_{\alpha'\beta'}\}\{\prod_j d\vec
\xi_j\}exp(-\beta(H_{eff}-Z^*ea_0\sum_i \vec \xi_{i}\cdot\vec E))}.
\label{strainthermexp}
\end{equation}
Differentiating Equation \ref{strainthermexp} with respect to $E_{\gamma}$, we
find the correlation function expression:
\begin{equation}
d_{\gamma\alpha\beta}=\beta(Z^*ea_0)
(<e_{\alpha\beta}\sum_j\xi_{j\gamma}>-N<e_{\alpha\beta}><\xi_{\gamma}>).
\label{piezocorrfcn}
\end{equation}

When the homogeneous strain appears only up to quadratic order in the effective
Hamiltonian, as is the case for the effective Hamiltonians available for
PbTiO$_3$\cite{ptwburg,ptprb}, BaTiO$_3$\cite{zvr} and KNbO$_3$\cite{krakauer},
the piezoelectric tensor can be reexpressed in terms of correlations of local
polar distortions. Because the integral over strain is Gaussian, completely
equivalent expressions for the thermal expectation values involving
$e_{\alpha\beta}$ can be obtained by making the following substitutions (and
cyclic permutations thereof):
\begin{equation}
e_{xx}\rightarrow-{1\over N}({(g_0+{g_1\over
3})\over(C_{11}+2C_{12})}\sum_i\vert\vec\xi_i\vert^2+{g_1\over
3(C_{11}-C_{12})}\sum_i(2\xi_{ix}^2-\xi_{iy}^2-\xi_{iz}^2))
\label{substitutionxx}
\end{equation}
\begin{equation}
e_{xy}\rightarrow-{1\over N}{g_2\over C_{44}}\sum_i\xi_{ix}\xi_{iy}
\label{substitutionxy}
\end{equation}
For the example of $d_{33}$, this yields
\begin{eqnarray}
d_{33}=-{\beta(Z^*ea_0) \over N}({(g_0+{g_1\over
3})\over(C_{11}+2C_{12})}\sum_{i,j}
(<\vert\vec\xi_i\vert^2\xi_{jz}>&-&<\vert\vec\xi_i\vert^2><\xi_{jz}>) \nonumber
\\
+{g_1\over 3(C_{11}-C_{12})}\sum_{i,j}
(2(<\xi_{ix}^2\xi_{jz}>&-&<\xi_{ix}^2><\xi_{jz}>) \nonumber \\
-(<\xi_{iy}^2\xi_{jz}>&-&<\xi_{iy}^2><\xi_{jz}>) \nonumber \\
-(<\xi_{iz}^2\xi_{jz}>&-&<\xi_{iz}^2><\xi_{jz}>))
\label{piezocorrfcna}
\end{eqnarray}

While Equation \ref{piezocorrfcn} proves more convenient for the actual
computation of $d_{\gamma\alpha\beta}$,  it is Equation \ref{piezocorrfcna}
that naturally leads to a microscopic interpretation of the soft-mode response,
as we will discuss further below.

\section*{$\epsilon_{\lowercase{zz}}$ and $\lowercase{d}_{33}$ for
P\lowercase{b}T\lowercase{i}O$_3$}

\begin{figure}[b!] 
\centerline{\epsfig{file=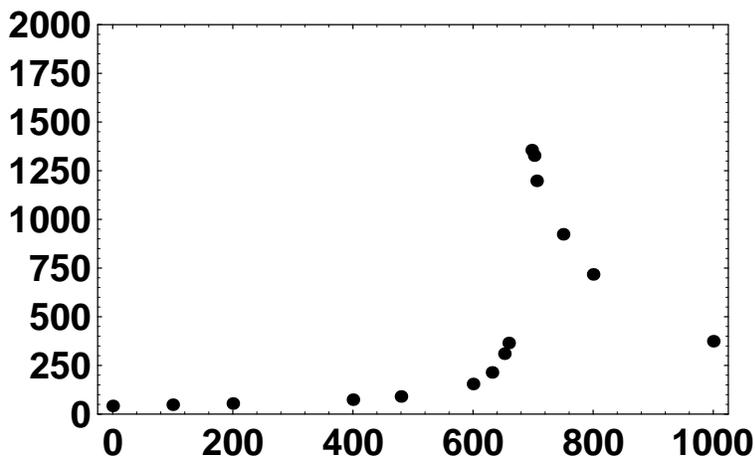}}
\vspace{10pt}
\caption{Dielectric response $\epsilon_{zz}$ as a function of temperature, in
K.}
\label{fig1}
\end{figure}

The results of the Monte Carlo calculations for $\epsilon_{zz}$ are shown in
Figure \ref{fig1}. At room temperature, the calculated value is 66, which
should be compared to the single-crystal measurement  $\epsilon_{zz}$ =
80\cite{li}. The agreement is surprisingly good, considering that the
calculation includes only the soft-mode contribution. The room temperature
value is already considerably larger than the zero-temperature calculated value
of 42. The overall temperature dependence is very similar to that observed
experimentally\cite{remeika,li}, with a slow increase with temperature below
the transition, a near-divergence both above and below T$_c$ which is cut off
by the first-order character of the cubic-tetragonal transition, and an
enhanced value of the dielectric response in the high-temperature cubic phase.

\begin{figure}[b!] 
\centerline{\epsfig{file=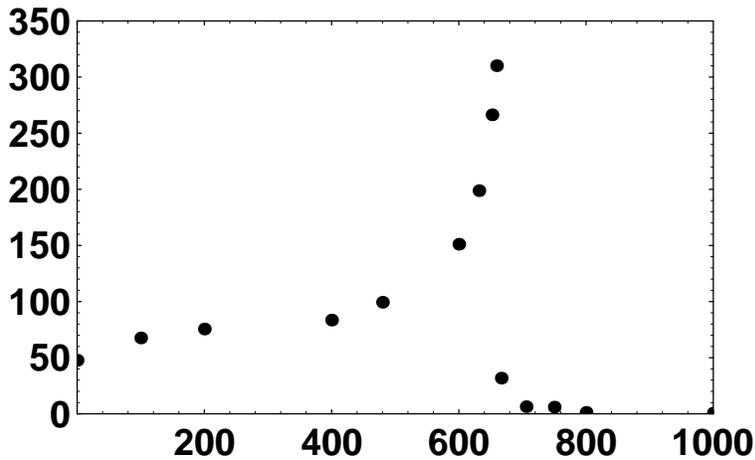}}
\vspace{10pt}
\caption{Piezoelectric response $d_{33}$ as a function of temperature, in K.}
\label{fig2}
\end{figure}

The results of the Monte Carlo calculations for $d_{33}$ are shown in Figure
\ref{fig2}. At room temperature, the calculated value is 76, which should be
compared to the single-crystal measurement  $d_{33}$ = 83.7\cite{li}. As for
the dielectric response, the agreement is surprisingly good, considering that
the calculation includes only the soft-mode contribution. The room temperature
value is already considerably larger than the zero-temperature calculated value
of 48. The overall temperature dependence is very similar to that observed
experimentally\cite{remeika,linew}, with a slow increase with temperature below
the transition, a near-divergence below T$_c$ which is cut off by the
first-order character of the cubic-tetragonal transition, and a drop to zero in
the high-temperature cubic phase, as required by symmetry.

These results can be compared with the closely analogous calculation of the
piezoelectric coefficients of BaTiO$_3$ (Figure 1 in \cite{vandgar}), noting
that for this system as well, remarkably good quantitative agreement with the
experimentally measured values is obtained.

\section*{Landau theory for the divergent responses}
The near-divergences observed in the results described above can be
already understood within a simple Landau theory which describes coupling of a
scalar polar mode $\xi$ with a one-component strain $e$ and electric field $E$:
\begin{equation}
F(\xi,e;T,E)=A\xi^2+B\xi^4-E\xi+ce^2+ge\xi^2
\end{equation}
where the temperature dependence enters through $A=A_0(T-T_c)$.
If we first eliminate $e$, which is accomplished by the substitution
$e\rightarrow{g\over c}\xi^2$, the following effective fourth-order free energy
results:
\begin{equation}
F(\xi,e;T,E)=A\xi^2+(B-{g^2\over 4c})\xi^4-E\xi
\end{equation}
Minimizing with respect to $\xi$ yields the following expressions for the
dielectric susceptibility
\begin{eqnarray}
\chi={d\xi_{min}\over dE}&=&{-1\over 4A}~~~(T<T_c)\nonumber \\
&=&{1\over 2A}~~~(T>T_c)
\end{eqnarray}
and piezoelectric coefficient
\begin{eqnarray}
d={de_{min}\over dE}&=&-{g\over 4Ac}\sqrt{{-A\over 2(B-{g^2\over
4c})}}=-{g\over c}\xi_{min}\chi~~~(T<T_c)\nonumber\\
&=&0~~~(T>T_c)
\end{eqnarray}
As observed in the full simulations, $d$ diverges as T$_c$ is approached from
below, while $\chi$ diverges on both the low and high temperature sides, and is
larger above T$_c$.
In PbTiO$_3$, these divergences are cut off due to the first-order character of
the transition, so that, for example, a Curie-Weiss fit to $\chi$ above T$_c$
yields a value of T$_0$ about 80 K below T$_c$.

\section*{Real-space decomposition of the correlation functions}
The correlation function expressions for the response functions allow the
interpretation of the responses in terms of the characteristic intersite
correlations between the local polar distortions.
Some of the correlation functions of interest have been previously calculated
in effective Hamiltonian studies of KNbO$_3$\cite{krakauer} and
BaTiO$_3$\cite{vanderbiltcorr}, allowing us to connect to and build upon the
insights derived in those investigations to develop a microscopic understanding
of the dielectric and piezoelectric responses.

For the dielectric response, the first term inside the parentheses of Equation
\ref{dicorrfcn} can be reorganized:
$${1\over N}\sum_i(<\xi_{i\alpha}\xi_{i\beta}>+\sum_{j,
nn}<\xi_{i\alpha}\xi_{j\beta}>+\sum_{j, nnn}<\xi_{i\alpha}\xi_{j\beta}>+...)$$
If correlations become infinite ranged, as at a second order transition, this
will scale like N, leading to a divergence in the thermodynamic limit.

For the piezoelectric tensor, we can similarly reorganize the correlation
functions which appear in Equation \ref{piezocorrfcna}, for example:
$${1\over N}\sum_i(<\xi_{i\alpha}\xi_{i\alpha}\xi_{i\gamma}>+\sum_{j,
nn}<\xi_{i\alpha}\xi_{i\alpha}\xi_{j\gamma}>+\sum_{j,
nnn}<\xi_{i\alpha}\xi_{i\alpha}\xi_{j\gamma}>+...)$$

The real-space correlation functions appearing in the expressions above are
computed using fast fourier transforms of $\vec \xi_i$ and
$\xi_{i\alpha}\xi_{i\beta}$.
For the dielectric susceptibility, for example, the real space expressions can
be written in terms of $\chi_{\alpha\beta}(\vec q)$ as follows (taking $\vec
R_j = \vec R_i + \vec d$):
$${1\over N}\sum_i <\xi_{i\alpha}\xi_{j\beta}>={1\over N}\sum_i{1\over
N^2}\sum_{\vec q, \vec q~'}<\xi_{\alpha}(\vec q)\xi_{\beta}(\vec
q~')>exp(-i\vec q \cdot \vec R_i-i\vec q~' \cdot (\vec R_i + \vec d))$$
$$={1\over N^2}\sum_{\vec q}<\xi_{\alpha}(\vec q)\xi_{\beta}(\vec
q)^*>exp(i\vec q \cdot \vec d)$$
It can be checked that this gives the correct relation to $\chi_{\alpha\beta}$
when $<\vec \xi> = 0$:
$$\chi_{\alpha\beta}= {\beta(Z^*ea_0)^2\over\Omega}\sum_{\vec d}{1\over
N}\sum_i<\xi_{i\alpha}\xi_{j\beta}>=\sum_{\vec d}{1\over N}\sum_{\vec
q}\chi_{\alpha\beta}(\vec q)exp(i\vec q \cdot \vec d)=\chi_{\alpha\beta}(\vec q
=0)$$

\begin{figure}[b!] 
\centerline{\epsfig{file=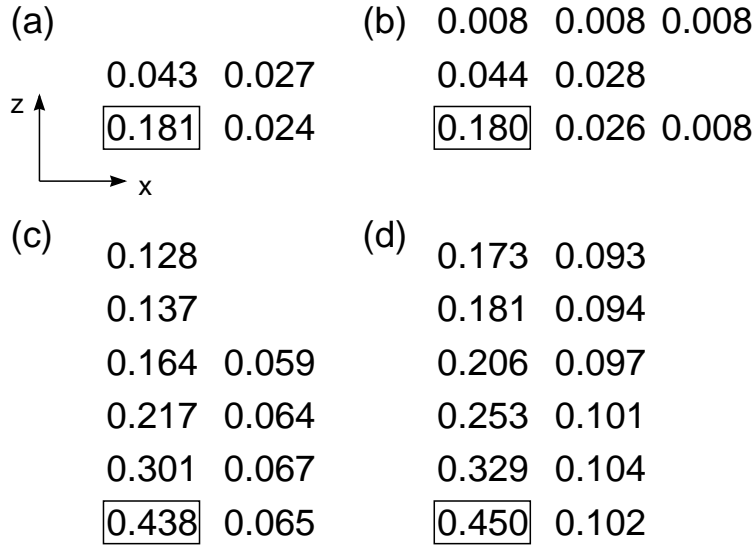,height=3in}}
\vspace{10pt}
\caption{Real-space decomposition in the $xz$ plane of the high-T dielectric
susceptibility correlation function $<\xi_{iz}\xi_{i+\vec d,z}>$ for (a)
PbTiO$_3$ at T = 706 K, (b) PbTiO$_3$ at T = 690 K, (c) BaTiO$_3$ at T = 343 K,
and (d) BaTiO$_3$ at T = 308 K. The numerical value in the box indicates the
on-site correlation$<\xi_{iz}\xi_{iz}>$, while the numbers above and to the
right of the box give the correlations for the corresponding values of $\vec
d$. All correlations have been normalized by dividing by the square of the
ground state value of $\xi_z$ (0.08$^2$ for PbTiO$_3$ and 0.03$^2$ for
BaTiO$_3$). Only the largest correlations are shown (above 0.0078 for PbTiO$_3$
and above 0.056 for BaTiO$_3$).}
\label{fig3}
\end{figure}

In Figure \ref{fig3},
we show the real-space decomposition in the $xz$ plane of the high-T dielectric
susceptibility correlation function $<\xi_{iz}\xi_{i+\vec d,z}>$ for PbTiO$_3$
at two temperatures above T$_c$. The correlation is rather isotropic, and the
range of the correlations increases as expected as the temperature decreases
towards T$_c$. For comparison, the same correlation function for BaTiO$_3$ at
two temperatures above the cubic-tetragonal transition temperature is also
shown in Figure \ref{fig3}. The highly anisotropic ``chain-like'' correlations
discussed in previous work \cite{krakauer,vanderbiltcorr} are clearly evident,
and the correlation length transverse to the chains increases as the
temperature decreases towards T$_c$. Similar behavior is expected for KNbO$_3$.
The qualitative difference in the nature of the correlations can be directly
attributed to the difference in the dispersion relation of the soft mode along
R. The implications of the different character of these correlations for the
dielectric behavior of PbTiO$_3$ and BaTiO$_3$ are currently under
investigation.

As can be seen from Equation \ref{piezocorrfcna}, the piezoelectric response
depends on third-order two-site correlations of the form
$<\xi_{i\alpha}^2\xi_{j\beta}>-<\xi_{i\alpha}^2><\xi_{j\beta}>$. These have not
to our knowledge been considered in previous studies. Calculations of these
real-space correlations for PbTiO$_3$ and BaTiO$_3$ above and below the
transition temperature are currently in progress.

\section*{Discussion and conclusions}

As already discussed, the use of an effective Hamiltonian for calculating the
dielectric and piezoelectric response involves a number of approximations.
The effects of thermal expansion and the neglect of other polar modes cannot be
incorporated without adding additional degrees of freedom to the model (unless
one is willing to accept semi-empirical input).
However, the low-order expansion around the cubic reference structure can be
improved, which in particular will yield better model results for the
properties of the PbTiO$_3$ tetragonal ground state \cite{umesh}.
This could involve both ${\cal H}_{eff}$ and $P$.
In the case of the polarization $P$, the next order terms involve including
strain dependence and $\xi$ dependence of the effective charges.
First-principles calculations\cite{wang96,ghosez} suggest that the former
corrections are rather small compared to the latter. Calculations for these
refinements for PbTiO$_3$ are currently in progress.

Since the effective Hamiltonian approach is limited to the description of
soft-mode contributions to the responses, we focus on the near-divergent
responses where these contributions strongly dominate.
Fortunately, this regime is, in any case, the one of greatest interest for
understanding the origin of and engineering large responses in a variety of
systems characterized by proximity to a lattice instability.
Further insight can be gained from the real-space decomposition of the
divergent responses based on their expressions in terms of correlation
functions.

Finally, these results for pure systems suggest an interpretation for the
origin of large responses in solid solutions. For such systems, there is a
distribution of local environments, and for a given temperature some subset of
lattice degrees of freedom will be marginally stable. This idea has been
investigated using first principles results in Pb$_{1-x}$Ge$_x$Te,  and more
details are presented in Refs.
\onlinecite{enhance,cockayne}.

\section*{Acknowledgments}
We thank R. E. Cohen, A. Garcia, R. M. Martin, D. Vanderbilt, and U. V.
Waghmare for helpful discussions. This work was supported by ONR grant
N00014-97-1-0047 and the Alfred P. Sloan Foundation, and was in part carried
out at the Aspen Center for Physics.


\begin{references}
\bibitem{transducer} J. M. Herbert,
{\it Ferroelectric Transducers and Sensors}, New York: Gordon and Breach, 1982.

\bibitem{remeika} J. P. Remeika and A. M. Glass, Mater. Res. Bull. {\bf 5}, 37
(1970).

\bibitem{jjap} K. Kakuta, T. Tsurumi and O. Fukunaga, Jpn. J. Appl. Phys. {\bf
34}, 5341 (1995).

\bibitem{linew} Z. Li, M. Grimsditch, C. M. Foster and S.-K. Chan, J. Phys.
Chem. Solids {\bf 57}, 1433 (1996).

\bibitem{btexpt} C. J. Johnson, Appl. Phys. Lett. {\bf 7}, 221 (1965).

\bibitem{PZT} F. Jona and G. Shirane,
{\it Ferroelectric Crystals}, New York: Macmillan, 1962.

\bibitem{rwwannier} K. M. Rabe and U. V. Waghmare, Phys. Rev.
{\bf B52}, 13236 (1995).

\bibitem{ptwburg} K. M. Rabe and U. V. Waghmare, J. Phys. Chem. Solids {\bf
57}, 1397 (1997).

\bibitem{ptprb} U. V. Waghmare and K. M. Rabe, Phys. Rev. {\bf B55},
6161 (1997).

\bibitem{zvr} W. Zhong, D. Vanderbilt, and K. M. Rabe, Phys. Rev. Lett. {\bf
73}, 1861 (1994); Phys. Rev. B {\bf 52}, 6301 (1995).

\bibitem{Nye} J. F. Nye,
{\it Physical Properties of Crystals}, Oxford: Clarendon, 1964.

\bibitem{wgarcia} A. Garcia and D. Vanderbilt, these proceedings.

\bibitem{krakauer} H. Krakauer, R. Yu, C.-Z. Wang, and C. LaSota,
  {\it Ferroelectrics}, in press; H. Krakauer, R. Yu, C.-Z. Wang, K. M. Rabe
and U. V. Waghmare, cond-mat/9710088.

\bibitem{li} Z. Li, M. Grimsditch, X. Xu and S.-K. Chan, Ferroelectrics {\bf
141}, 313 (1993).

\bibitem{vandgar} A. Garcia and D. Vanderbilt, cond-mat/9801177.

\bibitem{vanderbiltcorr} J. Padilla, W. Zhong and D. Vanderbilt,
mtrl-th/9509005.

\bibitem{wang96} C.-Z. Wang, R. Yu, and H. Krakauer, Phys. Rev. {\bf B54},
11161 (1996).

\bibitem{ghosez} Ph. Ghosez, X. Gonze, Ph. Lambin and J.-P. Michenaud, Phys.
Rev. {\bf B51}, R6765 (1995).

\bibitem{umesh} U. V. Waghmare, unpublished.

\bibitem{enhance} E. Cockayne and K. M. Rabe, cond-mat/9712232.

\bibitem{cockayne} E. Cockayne and K. M. Rabe, these proceedings.
\end{references}
\end{document}